\documentclass[a4paper]{article}

\usepackage{INTERSPEECH2022}

\title{Relating the fundamental frequency of speech with EEG using a dilated convolutional network}
\name{Corentin Puffay$^{1,2}$, Jana Van Canneyt$^1$, Jonas Vanthornhout$^1$, Hugo Van hamme$^2$, Tom Francart$^1$}
\address{
  $^1$KU Leuven, Dept. Neurosciences, ExpORL, Leuven, Belgium\\
  $^2$KU Leuven, Dept. of Electrical engineering (ESAT), PSI, Leuven, Belgium}
\email{corentin.puffay@kuleuven.be, jonas.vanthornhout@kuleuven.be, jana.vancanneyt@kuleuven.be,  hugo.vanhamme@kuleuven.be, tom.francart@kuleuven.be}

\begin{document}

\maketitle
\begin{abstract}
To investigate how speech is processed in the brain, we can model the relation between features of a natural speech signal and the corresponding recorded electroencephalogram (EEG). Usually, linear models are used in regression tasks. Either EEG is predicted, or speech is reconstructed, and the correlation between predicted and actual signal is used to measure the brain's decoding ability. However, given the nonlinear nature of the brain, the modeling ability of linear models is limited. Recent studies introduced nonlinear models to relate the speech envelope to EEG. We set out to include other features of speech that are not coded in the envelope, notably the fundamental frequency of the voice (f0). F0 is a higher-frequency feature primarily coded at the brainstem to midbrain level. We present a dilated-convolutional model to provide evidence of neural tracking of the f0. We show that a combination of f0 and the speech envelope improves the performance of a state-of-the-art envelope-based model. This suggests the dilated-convolutional model can extract non-redundant information from both f0 and the envelope. We also show the ability of the dilated-convolutional model to generalize to subjects not included during training. This latter finding will accelerate f0-based hearing diagnosis.\\
\end{abstract}

\noindent\textbf{Index Terms}: EEG decoding, speech, auditory system, fundamental frequency, envelope

\section{Introduction}

The human brain responds and time-locks to sound picked up by the ears. Electroencephalography (EEG) enables measuring this time-locking non-invasively. 
Traditionally, discrete stimuli are repeated several times and the recorded EEG signal is averaged out to obtain the brain response to a speech stimulus. This method cannot be used in real-life situations during which people listen to speech as the signal is long,  continuous and not often repetitive. Instead of averaging, we can model the mapping between a stimulus feature and the neural response, as described in \cite{Ding2012, Crosse2016}. Linear regression models are used either to reconstruct speech from EEG (backward models) or to predict EEG from speech (forward models) \cite{Crosse2016}. The correlation between the reconstructed (or predicted) speech (or EEG) and the ground truth is computed to measure what we will call neural tracking \cite{Ding2012, Crosse2016, Vanthornhout2018, DiLiberto2015}. Linear models can estimate neural tracking of different speech feature categories such as acoustics (i.e speech features not containing language-related information) and linguistics. For example, the speech envelope and the spectrogram were used \cite{Lesenfants2019}. In addition, neural tracking of the fundamental frequency of the voice (f0) has been shown when the speech stimuli were male voices \cite{VanCanneyt2021}. 
Neural tracking can be estimated with a regression as well as a classification paradigm \cite{DeCheveigne2021}. Here we will use a classification task: a match-mismatch task which involves a model deciding whether a brain signal segment matches with the stimulus that evoked it or with an mismatched stimulus segment.\\
Recent advances of deep learning in Automatic Speech Recognition (ASR) encourage the use of deep neural networks for EEG decoding. An LSTM-based model \cite{Monesi2020} as well as a dilated-convolution-based model \cite{Accou2021ModelingTR} notably outperformed linear models on the match-mismatch classification task with the speech envelope as a feature.\\
In this study, we present a dilated-convolutional neural network to measure f0-tracking in male-narrated stories from distinct speakers using the match-mismatch task.
In a second part, we integrate both f0 and the speech envelope in the model as two separate input streams that fuse in later layers. We show that using the combination of both features improves performance on the match-mismatch task.
Linear models are usually trained in a subject-dependent way, i.e., for each subject, a separate model is trained on training data of that subject. For linear models this yields better performance than training on (multiple) other subjects. For the envelope-based deep learning models \cite{Accou2021ModelingTR,Monesi2020}, it has been shown that \emph{better} performance can be achieved by training on a relatively large number of other subjects. We investigate whether the same holds for our f0 model. 

\section{Methods}

\subsection{Data collection}
For this dataset, 60 normal-hearing native-Flemish-speaking subjects between 18 and 30 years old were recruited. This study was approved by the Medical Ethics Committee UZ KU Leuven/Research (KU Leuven, Belgium) with reference S57102 and all subjects provided informed consent. In our protocol, we present natural running speech (stories) to subjects without background noise and record the EEG signal simultaneously. All subjects went through screening for normal hearing with pure tone audiometry and the Flemish Matrix test \cite{Luts2014DevelopmentAN}. 
All the subjects listened to a set of 10 unique stories of roughly the same duration (on average 14 minutes 30 seconds). The presentation order of the stories was randomized for each subject. The stories were presented binaurally at 62 dBA with shielded ER-3A insert phones (Etymotic, Elk Grove Village, Illinois, United States). Their only task was to answer a comprehension question after listening to each story to ensure they paid attention. EEG data was recorded using a 64-channel Active-Two EEG system (BioSemi, Amsterdam, Netherlands)  at 8192~Hz sampling rate. The stimuli were presented using the APEX 4 software platform developed at ExpORL \cite{APEX}. The experiments took place in a electromagnetically shielded and soundproofed cabin.

\subsection{Feature extraction and pre-processing}

\subsubsection{Speech envelope}
The envelope of the speech stimulus was computed using the powerlaw sub-band method from \cite{Biesmans2017AuditoryInspiredSE}. This method uses a gammatone filter bank \cite{Patterson1995TimedomainMO, SoendergaardMajdak2013} to extract each N=28 frequency sub-bands of the speech signal into a vector $s(t)$. For each sub-band, $|s(t)|^{0.6}$ is calculated. Finally, these sub-bands were averaged to obtain the speech envelope feature.

\subsubsection{F0}
The f0 feature was obtained by band-pass filtering the stimulus with a Chebyshev type-II filter with 80~dB attenuation at 10\% outside the pass-band and a pass band ripple of 1 dB \cite{Etard2019, VanCanneyt2021}. The filter cut-offs were based on the distribution of the f0 of our 10 stories defined in \cite{VanCanneyt2021}: the stimulus was filtered between 75 and 175~Hz for male-narrated stories and between 120 and 300~Hz for female-narrated stories. 

\subsubsection{Pre-processing}

Then EEG was first downsampled to 1024~Hz. A multi-channel Wiener filter \cite{Somers2018} was then used to remove artefacts and re-referencing was performed to the average of all electrodes. Then the EEG was filtered identically to the speech features. In the case of the envelope, the speech and the EEG signal were band-passed filtered between 1 and 40~Hz. In the case of f0, the EEG signal was band-pass filtered between 75 and 175~Hz for male-narrated stories and between 120 and 300~Hz for female-narrated stories. One should also note that f0 is not present in unvoiced and silent sections, it is a common practice to set them to 0. We will not apply such masking to avoid removing relevant information from the high-frequency band. The speech envelope and corresponding EEG were downsampled to 64~Hz, whereas it was kept to 1024~Hz for f0 and its corresponding EEG. Furthermore, we divided the data of each subject into training, validation and testing sets. The first and last 40\% of the recording segment were used for training, the remaining 20\% were divided into 10\% for either validation or testing. 

\subsection{Match-mismatch classification task}
\begin{figure}[htbp!]
    \centering
    \includegraphics[width=0.5\textwidth]{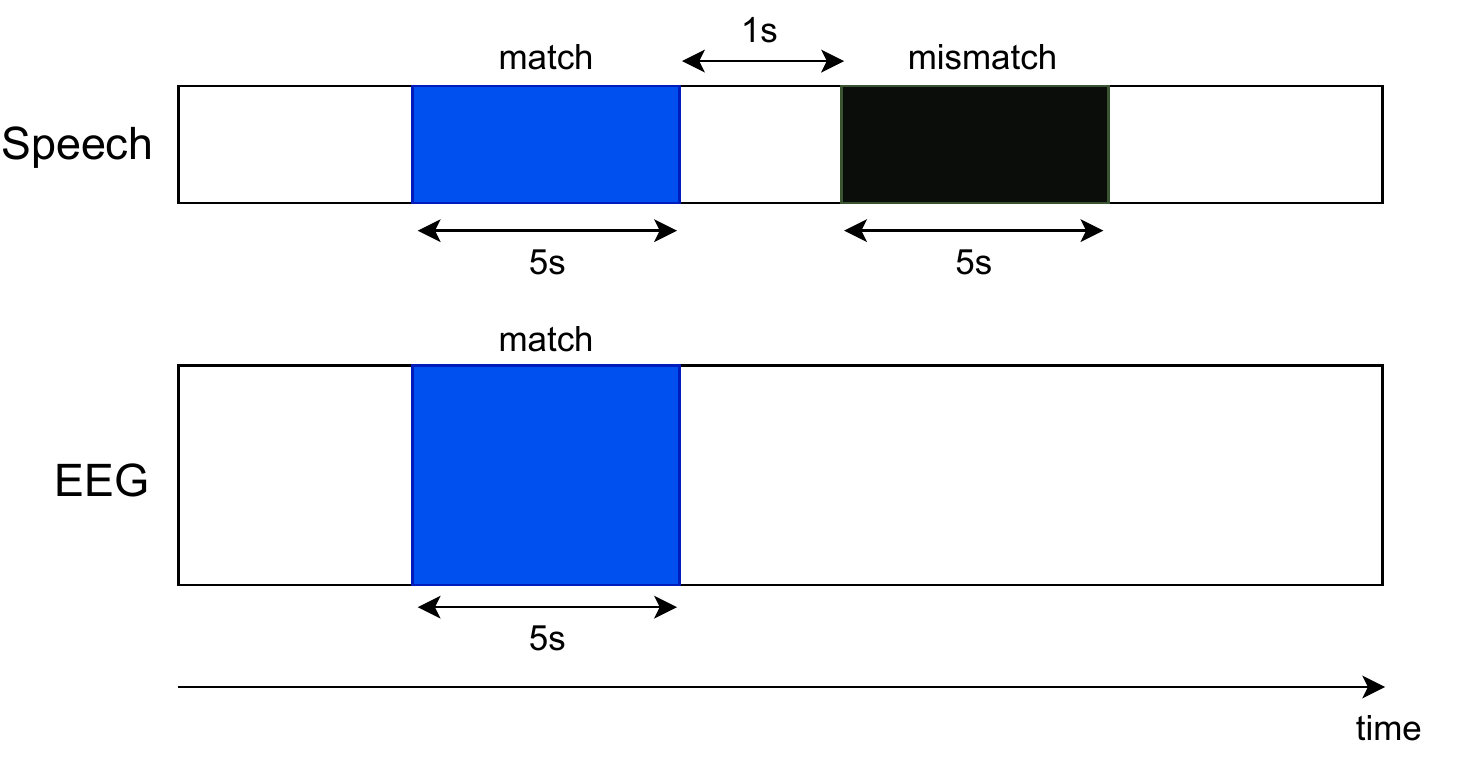}
    \caption{\textbf{Match-mismatch classification task}. The match-mismatch task is a binary classification paradigm which consists in associating the herewith blue EEG and speech segments. The matched speech segment is synchronized with the EEG while the mismatched speech is the segment occurring 1 second after the end of the matched segment. The figure depicts segments of 5s, however different duration are picked throughout our analyses. } 
    \label{fig:MM}
\end{figure}

The performance on a match-mismatch classification task is used in this study as a measure of neural tracking of different speech features. This paradigm is depicted on Figure~\ref{fig:MM}. The model is trained to associate the EEG segment with the matching speech segment. The matched speech segment is synchronized with the EEG while the mismatched speech is the segment occurring 1 second after the end of the matched segment. These segments are of fixed length (i.e 5~s or 2~s in this study).
This task is supervised as the match and mismatch segments are labeled. The evaluation metric is the classification accuracy.

\subsection{Models}

All the described models were created in Tensorflow (2.3.0) \cite{tensorflow} with the Keras API \cite{chollet2015keras}.

\subsubsection{Single feature model architecture}
\begin{figure}[htbp!]
    \centering
     \hspace{-0.5cm}
    \includegraphics[width=0.50\textwidth]{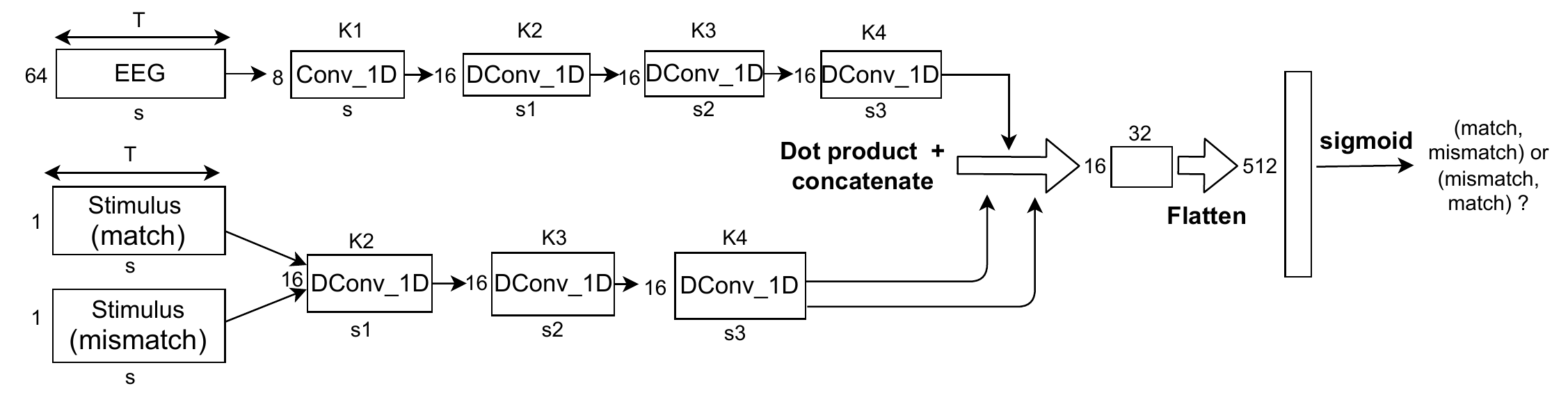}
   
    \caption{\textbf{Single-speech-feature  dilated-convolutional model.}} 
    \label{fig:dilation_single_feat}
\end{figure}

The first model we used is a single-speech-feature model (see Figure~\ref{fig:dilation_single_feat}). It takes one EEG segment (EEG) and two speech segments (labeled as match and mismatch) as inputs and classifies which of the two matches the EEG segment. These 3 segments are of same duration T and are provided in batches of size 64. To avoid any bias regarding label position, the mismatched segment is alternately presented to each of the speech envelope inputs. This means that each segment of EEG is presented twice to the network.\\
The EEG stream consists of one spatial filter (i.e. a convolution with a kernel size of 1) and three dilated convolutions with a maximum kernel size (see K4 on Figure~\ref{fig:dilation_single_feat}) defined per speech feature (422~ms in the speech envelope case, and 36~ms for f0). We choose these receptive fields as these are the maximum significant response delays observed in linear models \cite{Crosse2016, VanCanneyt2021}. The speech stream only contains the three dilated convolutions as no spatial filtering is needed (the envelope is 1-dimensional). For both EEG and speech streams, a rectified linear unit (ReLU) nonlinearity is applied after each dilated convolution.
Cosine similarities between the EEG and each speech embedding are computed (i.e. dot product between each EEG and speech row combinations divided by the L2-norm of both vectors, resulting in a 16x16 matrix for both match and mismatch stimuli). These two matrices are concatenated into a single 16x32 matrix further flattened into a 512x1 vector. The latter is provided to a fully-connected layer with a sigmoid as activation function to solve the binary classification task (i.e match or mismatch).\\
The model is trained in an end-to-end fashion for 50 epochs using early stopping with stochastic gradient descent (Adam optimizer, learning rate=0.001). The loss function used is the binary cross entropy.
Except the kernel sizes of dilated convolutions, the hyper-parameters are the same in the single-feature f0 and envelope models. The sampling rate of the f0 and the speech envelope are also different (1024~Hz and 64~Hz respectively), thus the time dimension of the inputs is different for the same segment length (e.g for 5~s, f0 segment length is 10240 samples whereas for the speech envelope it is 640 samples).\\
We will refer to the single-feature model as \textit{env} when using the speech envelope and \textit{f0} when using the fundamental frequency of the voice.

\subsubsection{N-speech-features model architecture}
\begin{figure}[htbp!]
    \centering
    \includegraphics[width=0.50\textwidth]{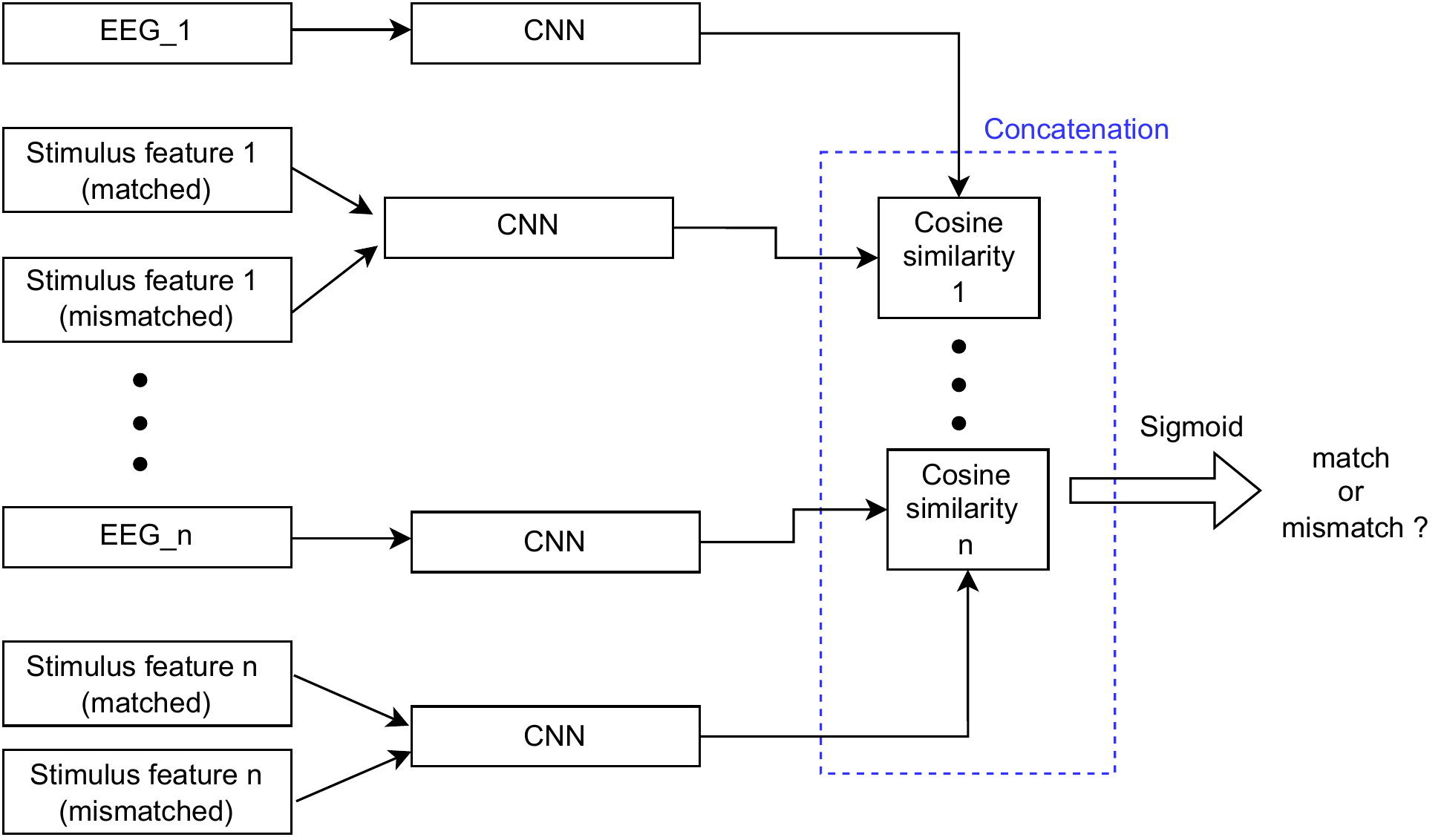}
    \caption{\textbf{N-speech-feature dilated-convolutional model} This is an extension of the previous model. There are N EEG streams corresponding to the N different speech features integrated in the model. For our specific case (N=2), Stimulus feature 1 is envelope and Stimulus feature 2 is f0.} 
    \label{fig:dilation_double_feat}
\end{figure}

The second model we used is a N-speech-feature model (see Figure~\ref{fig:dilation_double_feat}). In our experiments, we integrated the speech envelope and f0 (thus, N=2). The model takes two synchronized EEG segments (EEG\_1 and EEG\_2)  as inputs for stimulus feature 1 (envelope) and 2 (f0). EEG\_1 and EEG\_2 have the corresponding sampling rate and are obtained with the filtering of their respective stimulus. The model classifies which stimulus feature out of each pair (envelope match, f0 match) or (envelope mismatch, f0 mismatch) matches with (EEG\_1, EEG\_2). The hyper-parameters and the model steps are the same as in the single-speech feature model until the cosine similarity matrix is obtained with the corresponding speech feature. The additional step is the concatenation of the two 16x32 cosine similarity matrices corresponding to feature 1 and feature 2.
One EEG stream per speech feature is chosen as they may have different sampling rates (here 1024~Hz and 64~Hz for f0 and envelope respectively). We will refer the two-feature model integrating the speech envelope and f0 as \textit{env+f0}.\\
In the following experiments, the segment duration T will either be 5 or 2~s. Reducing this parameter makes the task more difficult for the network and will produce a drop in the classification accuracy.

\subsection{Subject-dependent vs subject-independent models}

We compared subject-dependent (SD) against subject-independent (SI) training for both our male-narrated stories (M1 and M2) using our single-feature (f0) model. We either used data from M1 only (\textit{M1}), M2 only (\textit{M2}), or both M1 and M2 (\textit{M1+M2}).
The SD paradigm consists of training and evaluating a model on each subject individually, whereas SI involves training on a set of subjects and evaluating on a different one. For the SD training condition, we trained one model per subject (20 in total) and kept the 80:10:10 ratio for training, validation and testing sets respectively. For the SI training condition, we used 40 and 20 distinct subjects (used for the SD condition) for training and testing respectively.

\section{Results}

\subsection{Single-feature f0 model}

\begin{figure}[htbp!]
    \centering
    \includegraphics[width=0.5\textwidth]{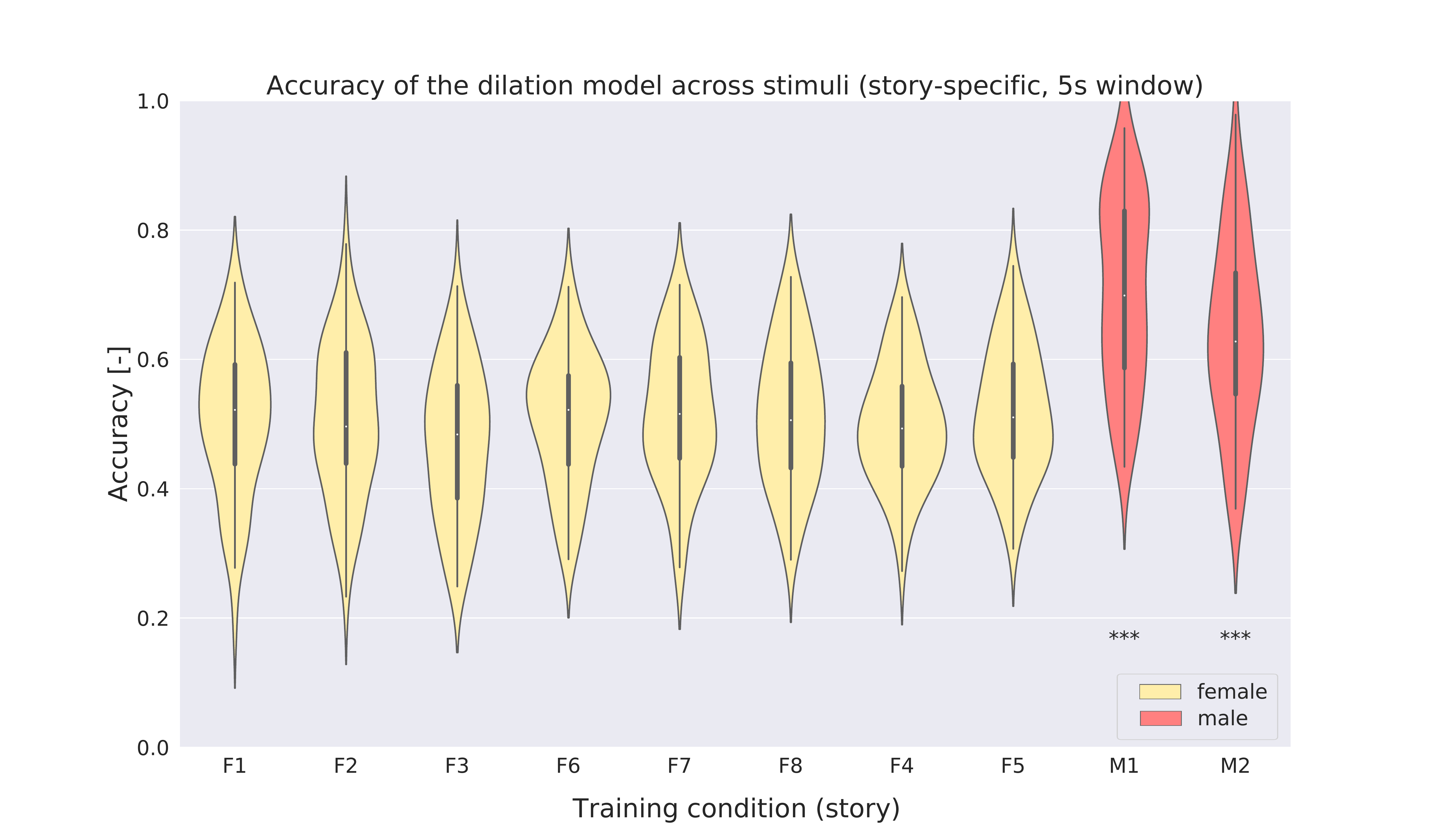}
    \caption{\textbf{Dilated-convolutional model (f0) accuracy across male- (M) and female-narrated (F) stories on the match-mismatch classification task for 5~s segments.} The digit following M or F is an index to denote the story number. Each point of these boxplots is the average performance across the test sets of 60 subjects. \textit{(*: p $<$0.05, **: p$<$0.01, ***: p$<$0.001)}} 
    \label{fig:f0_across_stories}
\end{figure}

Figure~\ref{fig:f0_across_stories} depicts the average accuracy per subject on the match-mismatch task (y-axis) for each of our 10 stories (x-axis). The distribution of the accuracies is represented by violin plots and it was tested for each story against chance level (0.5).
We observe that only for male-narrated stories (M1 and M2, in red) the model performs significantly above chance level (Wilcoxon signed-rank test against chance level, p$<$ 0.001). 
\noindent In that what follows we will only use male-narrated stories.

\subsection{Two-feature model}

\begin{figure}[h]
    \centering
    \includegraphics[width=0.50\textwidth]{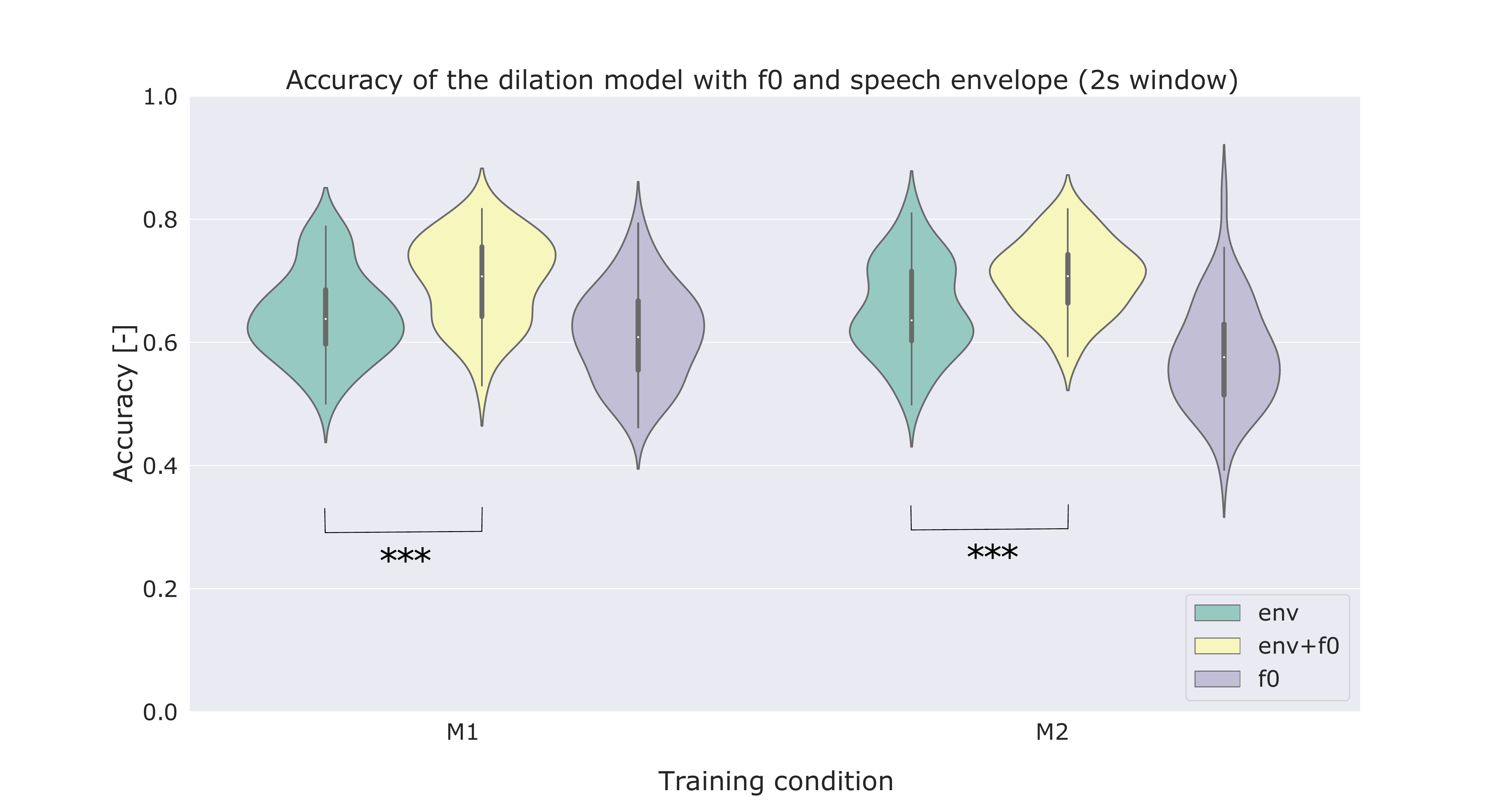}
    \caption{\textbf{Dilated-convolutional model accuracy when integrating envelope and f0 on male-narrated stories on the match-mismatch classification task for 2s segments.} Env or f0 refers to the single-feature model with the speech envelope or f0 while env+f0 refers to the two-feature model. Stories M1 and M2 are depicted here as they previously showed f0-tracking. \textit{(*: p$<$0.05, **: p$<$0.01, ***: p$<$0.001)}} 
    \label{fig:f0_env}
\end{figure}

In Figure~\ref{fig:f0_env}, the results with single-feature envelope and f0 models are compared with the two-feature model. We observe that when integrating f0 in addition to the envelope speech feature, the accuracy significantly increases compared to the single-feature envelope model for both male-narrated stories (Wilcoxon signed-rank test, p$<$0.001). When choosing longer input segments (e.g., 5~s), the median accuracy obtained on the match-mismatch task is around 0.9 which makes it difficult to compare distributions as lots of subjects approach the maximum accuracy (1). We therefore chose 2~s input segment length to avoid these ceiling effects, which could bias our conclusions.

\begin{figure}
    \includegraphics[width=0.50\textwidth]{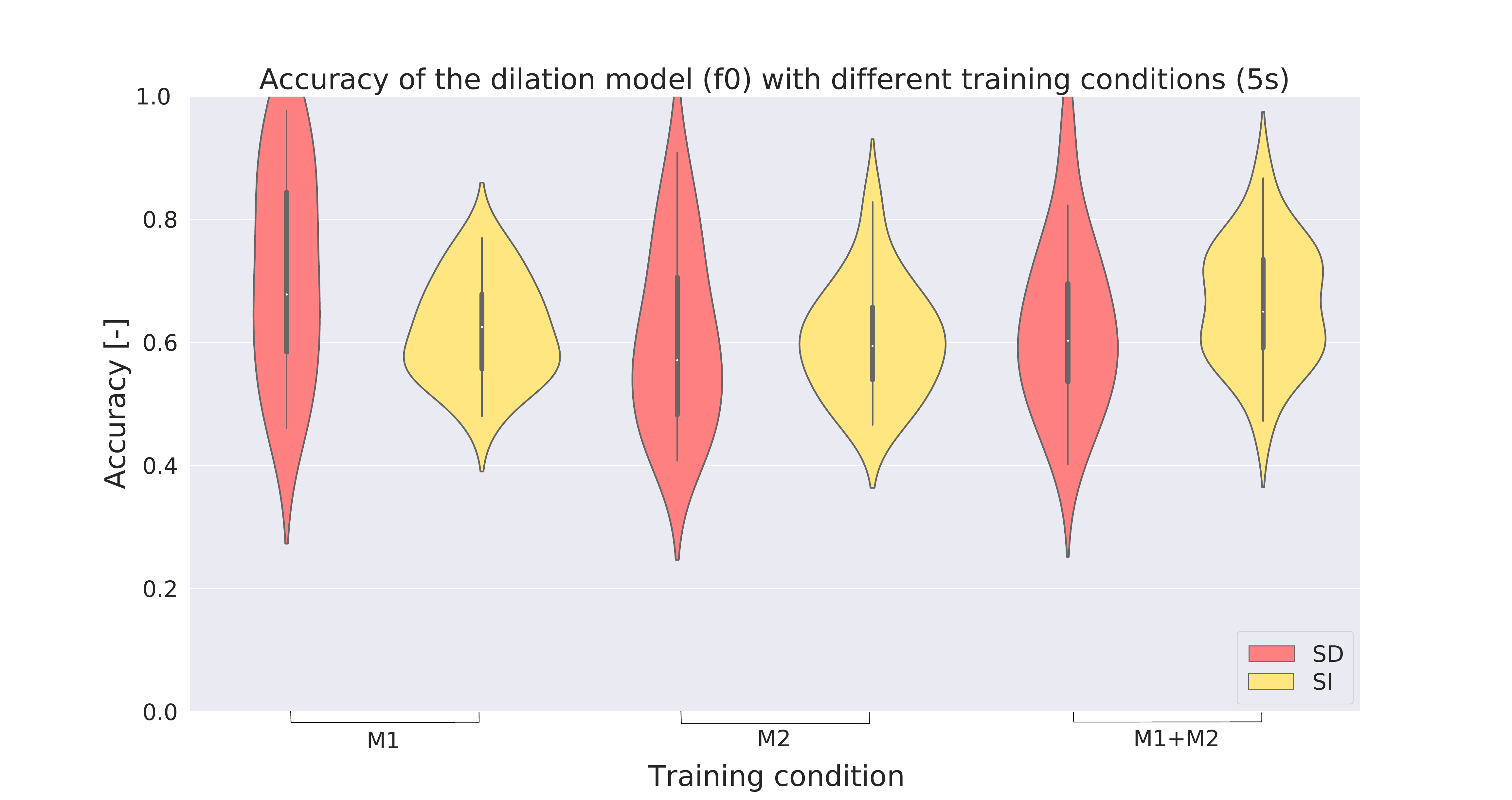}
    \caption{\textbf{Dilated-convolutional model (f0) accuracy under different training conditions.} The model was trained with 6 different conditions: subject-dependent for M1 or M2 (SD M1 or SD M2), subject-independent for M1 or M2 (SI M1 or SI M2), subject-dependent for M1 and M2 (SD M1+M2), subject-independent for M1 and M2 (SI M1+M2). \textit{(*: p$<$0.05, **: p$<$0.01, ***: p$<$0.001)}} 
    \label{fig:SS_SI}
\end{figure}

\subsection{Subject-independent and subject-dependent models}

The results are depicted in Figure~\ref{fig:SS_SI} and the different conditions (SD vs SI)  were compared pair-wise using a Wilcoxon signed-rank test. No significant increase was observed between SD and SI training conditions for every story combination (M1 only, M2 only or M1 and M2).


\section{Discussion}


We demonstrated that tracking of f0 by the brain can be measured with the dilated-convolutional model. We confirmed that f0-tracking could only be measured in male-narrated stories, which corroborates recent findings with linear methods \cite{VanCanneyt2021}. The inability to measure f0-tracking in female-narrated stories is problematic as it restricts our analysis with this feature. A hypothesis to explain this failure is that the skull acts as a low-pass filter \cite{Ramesh1998}, preventing the EEG electrodes from capturing high frequencies. The female-narrated voices of our stories oscillate in a higher frequency range than male-narrated voices, which might explain their absence in the EEG signal. Another explanation could be a female f0 has more variability (i.e., its rate of change is higher) compared to male f0, which makes it more challenging to track by the brain \cite{VanCanneyt2021}. \\
We showed that performance on the match-mismatch task increases when combining two speech features such as the speech envelope and f0. This result suggests that additional information carried by f0 is not present in the speech envelope and is also related to the EEG segment.\\
Finally, we demonstrated that the single-feature (f0) model does not significantly differ in performance with a SI or an SD training on a 5~s-segment classification task. This indicates the model can learn general patterns in the brain response to f0 and perform well on subjects not used in the training set. When including both M1 and M2, although not significantly, the SI model performs slightly better. Observing f0-tracking when including multiple stories into our dataset suggests the ability of the SI model to generalize to different male-narrated stories. F0 is primarily coded at the brainstem to midbrain level \cite{Micheyl2013}, which possibly mitigates its subject-specific component. Similar lower-brain-level responses could explain why the model can generalize to unseen subjects.\\
The dilated-convolutional model can measure f0- or envelope-tracking and both combined. This could provide the basis for a clinical tool to assess how well a person can understand speech. An objective measure of speech understanding could be derived from the match-mismatch accuracy following up on a recent study using the envelope \cite{Accou2021PredictingSI}. This study could, for instance, be extended using f0 to predict the speech reception threshold (SRT).\\
Another application is to model the auditory system. Every speech feature level (e.g., acoustic, lexical, or linguistic) is processed by different brain areas. For example, f0 has a significant brainstem response component while the envelope response is primarily cortical. The dilated-convolutional model processes features in separate branches like the auditory brain processing. We aim at integrating more speech features to have a complete deep-learning-based auditory model.\\

\section{Conclusions}

We aimed to use a dilated-convolutional neural network to model the relationship between EEG and single or multiple acoustic features.
We draw three main conclusions.
Firstly, f0-tracking is measured only in male-narrated stories, which corroborates previous findings with linear methods. Secondly, the combination of f0 and the speech envelope led to a higher classification accuracy on the match-mismatch task, suggesting an added value of f0 over the envelope and inversely. Finally, no significant difference in the single-feature f0 model's performance was found between the subject-dependent and subject-independent training conditions. The latter finding indicates that subject-specific training data is not required in practical applications. Future work will aim at other speech features integration. The more speech feature levels we can track, the more information we get about how the human brain perceives speech. We could, for instance, consider adding linguistic features (i.e., speech features providing information about grammar and correct syntax of a sentence).

\section{Acknowledgements}

The authors thank all the subjects for the recordings as well as Wendy Verheijen, Bernd Accou, Kyara Cloes, Amelie Algoet, Jolien Smeulders, Lore Kerkhofs, Sara Peeters, Merel Dillen, Ilham Gamgami, Amber Verhoeven, Lies Bollens, Vitor Vasconcelos and Amber Aerts for their help with data collection.
Funding was provided by the KU Leuven Special Research Fund C24/18/099 (C2 project to Tom Francart and Hugo Van hamme), FWO research project G0D6720N, and an FWO postdoctoral fellowship to Jonas Vanthornhout (1290821N).

\bibliographystyle{IEEEtran}

\bibliography{references.bib}


\end{document}